\documentclass[journal]{IEEEtran}
\usepackage{amsmath,amsfonts}
\usepackage{multirow}%
\usepackage{algorithm}%
\usepackage{algorithmicx}%
\usepackage{algpseudocode}%
\usepackage{array}
\usepackage[caption=false,font=footnotesize,labelfont=rm,textfont=rm]{subfig}
\usepackage{textcomp}
\usepackage{stfloats}
\usepackage{url}
\usepackage{verbatim}
\usepackage{graphicx}
\usepackage{listings}%
\usepackage{cite}
\usepackage{booktabs}
\usepackage{subfig}
\usepackage{threeparttable}
\usepackage{makecell}
\usepackage{doi}

\begin{document}

\title{\huge MAMCA - Optimal on Accuracy and Efficiency for Automatic Modulation Classification with Extended Signal Length}
\author{Yezhuo Zhang, Zinan Zhou, Yichao Cao, Guangyu Li and Xuanpeng Li
\thanks{This work has been submitted to the IEEE for possible publication. Copyright may be transferred without notice, after which this version may no longer be accessible. \textit{(Corresponding author: Xuanpeng Li.)}}
\thanks{Yezhuo Zhang, Zinan Zhou, Yichao Cao and Xuanpeng Li are with the School of Instrument Science and Engineering, Southeast University, Nanjing, 210096, Jiangsu, China (e-mail: li\_xuanpeng@seu.edu.cn). Guangyu Li is with the School of Computer Science and Engineering, University of Science and Technology, Nanjing, 210094, Jiangsu, China.}
}

\markboth{Journal of \LaTeX\ Class Files,~Vol.~14, No.~8, August~2021}%
{Shell \MakeLowercase{\underline{et al.}}: A Sample Article Using IEEEtran.cls for IEEE Journals}

\maketitle
\begin{abstract}
  With the rapid growth of the Internet of Things ecosystem, Automatic Modulation Classification (AMC) has become increasingly paramount.  However, extended signal lengths offer a bounty of information, yet impede the model's adaptability, introduce more noise interference, extend the training and inference time, and increase storage overhead. To bridge the gap between these requisites, we propose a novel AMC framework, designated as the Mamba-based Automatic Modulation ClassificAtion (MAMCA). Our method adeptly addresses the accuracy and efficiency requirements for long-sequence AMC. Specifically, we introduce the Selective State Space Model as the backbone, enhancing the model efficiency by reducing the dimensions of the state matrices and diminishing the frequency of information exchange across GPU memories. We design a denoising-capable unit to elevate the network's performance under low signal-to-noise radio. Rigorous experimental evaluations on the publicly available dataset RML2016.10, along with our synthetic dataset within multiple quadrature amplitude modulations and lengths, affirm that MAMCA delivers superior recognition accuracy while necessitating minimal computational time and memory occupancy. Codes are available on {\textit{\underline{https://github.com/ZhangYezhuo/MAMCA}}}.
\end{abstract}

\begin{IEEEkeywords}
Automatic modulation classification, selective state space model, cognitive radio, long sequence.
\end{IEEEkeywords}

\section{Introduction}\label{sec1}
\IEEEPARstart{I}{n} the realm of modern communication systems, Automatic Modulation Classification (AMC) has emerged offering essential functionality for numerous applications like spectrum monitoring, electronic warfare, and adaptive communication systems \cite{peng_survey_2022}. The drive toward more intelligent and agile communication systems has accelerated the research of AMC methodologies particularly in deep neural networks \cite{ma_automatic_2023}.

In the exploration of deep learning-driven AMC, researchers have adapted models such as Convolutional Neural Network (CNN), Long Short Term Memory (LSTM), Gate Recurrent Unit (GRU) and Transformer. Due to the low Signal-to-Noise Ratio (SNR), simple models may lose analytical capability, where recognition accuracy becomes a focal metric. Researchers combined complex modules to enhance the capacity of the classifier facing various noise inference yet increased the computational time and memory occupancy \cite{kumar_automatic_2024, yunhao_ConvLSTMAE_2022, li_automatic_2023}.

Concurrently, with the advancement of 5G/6G, especially the increasing Internet-of-Things (IoT) devices, the computing power at the terminal is limited while timeliness is emphasized \cite{guo_ultra_2024}. This has made efficiency as an equally critical metric. Prior studies have attained high timeliness and low memory occupancy by substantially reducing model complexity, and yet this has inevitably compromised accuracy \cite{guo_ultra_2024,aer_compressed_2022,liu_towards_2023}. Accuracy and efficiency present an antagonistic relationship where powerful models require higher computational costs, and lightweight models struggle to cope with low SNR and complex modulation changes.

\begin{figure}[t]
  \centering
  \subfloat[The effect of signal length on accuracy and efficiency in AMC task.]{\includegraphics[width=1\columnwidth]{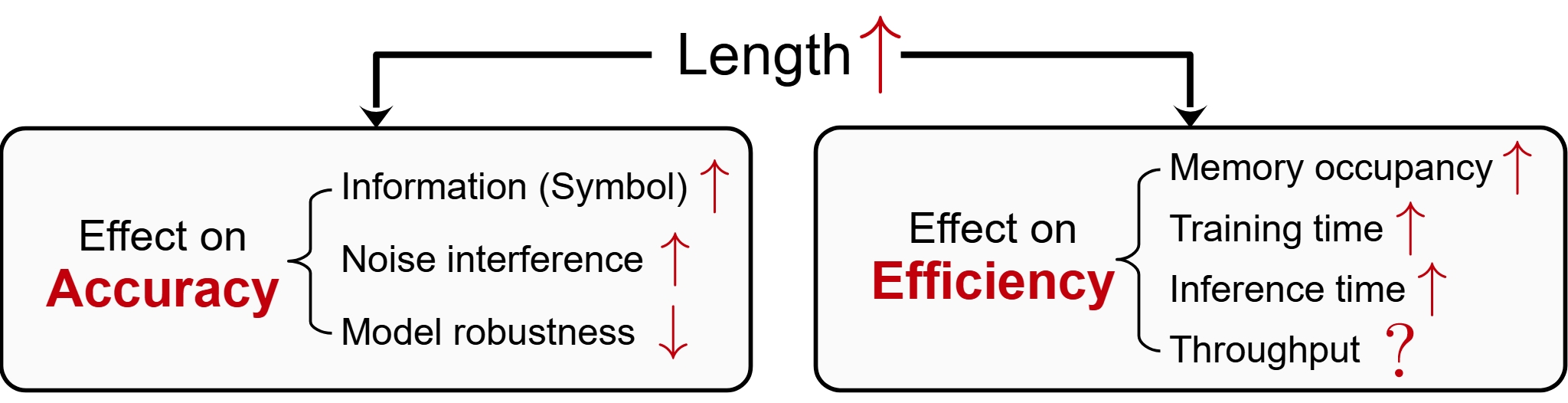}\label{length_effects}}\\
  \subfloat[A model that achieves Pareto optimal in terms of accuracy and efficiency.]{\includegraphics[width=1\columnwidth]{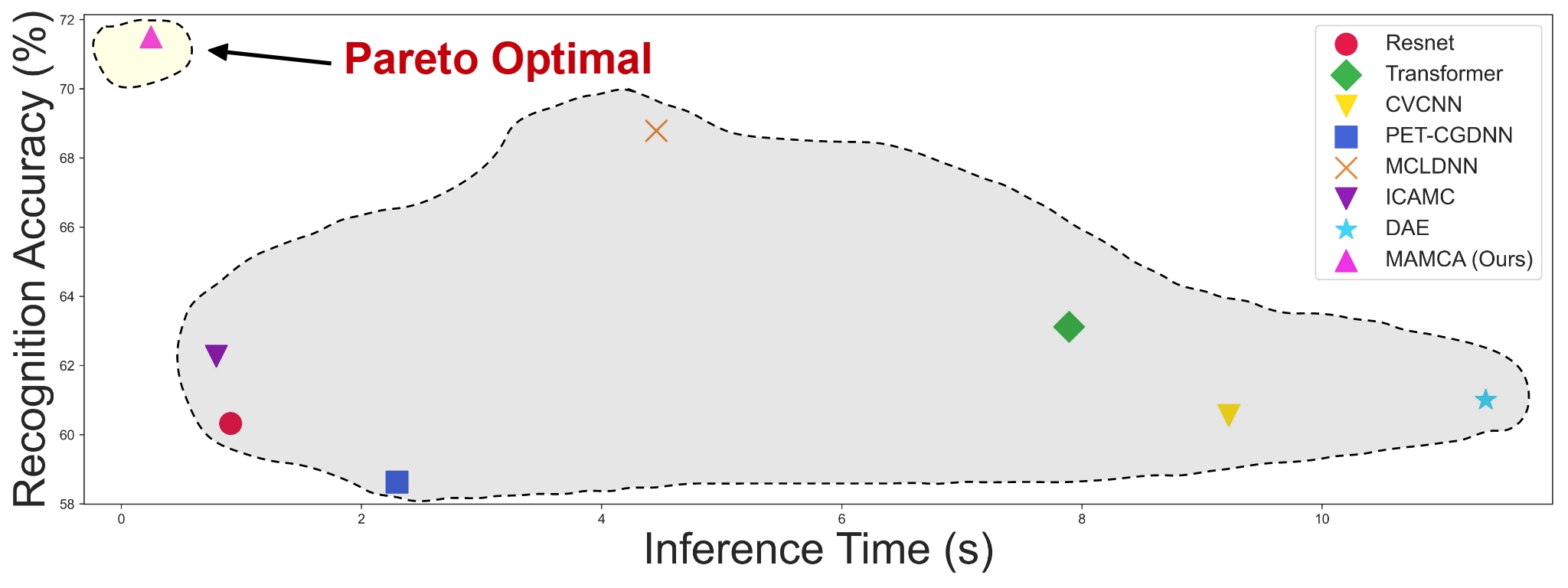}\label{optimal}}
  \caption{The impact of extended length, as well as the performance of models in addressing this impact.}\label{fig1}
  \vspace{-.4cm}
\end{figure}



Furthermore, an often overlooked challenge is that public datasets typically encompass relatively short sampling lengths, such as those exemplified by RML2016.10 \cite{Timothy_radio_2016}. The quantity of symbols within a sequence is constrained and may fall short of the actual counts prevalent in genuine modulation schemes. Nonetheless, current models seldom validate robustness with longer sequences, thus increased sequence length complicates the learning process and diminishes accuracy, while simultaneously escalating memories and reduces timeliness. As depicted in Fig. \ref{length_effects} with the effects of extended signal length, we aspire to develop a model as depicted in Fig. \ref{optimal} which can analyze longer sequences at once, and while increasing the sampling length, it can guarantee the recognition accuracy while ensuring efficiency on time and memory. 

An emerging model that meets these stringent requirements proposed, denoted as Mamba \cite{albert_mamba_2023}, which is a novel selective State Space Model (SSM) that elevates sequence modeling ability through input-driven parameter functions. This dynamic enables selective information propagation, benefiting tasks with long-range dependencies. Additionally, Mamba utilizes a hardware-optimized recurrent algorithm that mitigates input-output operations within GPU memory hierarchies, accelerating processing against prior models. 

In this letter, we propose the Mamba-based Automatic Modulation ClassificAtion (MAMCA) method. We introduce a denoising-capable convolutional unit to improve the model's adaptability on different noise interferences, incorporating Mamba to achieve accurate and efficient AMC. The contributions of this letter are as follows: 

\begin{itemize}
  \item We first introduce Selective SSM in AMC and have contributed to the community with code source-open. This initiative has established a novel benchmark in the AMC field, catalyzing further intensive research.
  \item We delved into the influence of signal sample length extension in AMC. The proposed MAMCA framework with denoising and time-dependence ability possesses greater robustness for longer sequence time-series.
  \item Our MAMCA framework demonstrates superlative performance. Relative to extant models, it consistently delivers superior accuracy, reduced computation time and memory occupancy, and stable performance across a spectrum of signal lengths and SNRs.
\end{itemize}

\section{Problem Statement and Formulation}\label{sec2}
In the formulation of the signal model, we consider a received signal within a wireless communication matrix with digital modulation delineated as: 
\begin{equation}
  \begin{aligned}
    r(t)=\mathcal{M}(s(t))*p(t)+n(t)
  \end{aligned}\label{eq1}
\end{equation}where $\mathcal{M}$ denotes the modulation operation applied to the transmitted symbol $s(t)$. $p(t)$ signifies the channel's impulse response, and $n(t)$ denotes additive white Gaussian noise. 

The analog representation is demodulated into orthogonal In-phase (I) and Quadrature-phase (Q) components, sampled to obtain digital signals as: 
\begin{equation}
  \begin{aligned}
  {r}(l)=\begin{bmatrix}\mathbf{I}(l)\\\mathbf{Q}(l)\end{bmatrix}=\begin{bmatrix}\mathbf{I}(0),\mathbf{I}(1),\ldots,\mathbf{I}(L-1)\\\mathbf{Q}(0),\mathbf{Q}(1),\ldots,\mathbf{Q}(L-1)\end{bmatrix}
  \end{aligned}\label{eq2}
\end{equation}

In the Automatic task, we collect a dataset $\mathcal{X}=\{x_i\}_{i=1}^N$ consisting of $N$ known modulation types. Our objective is to find a hypothesis space $h$ such that the model's prediction $\mathcal{Y}=\{y_i\}_{i=1}^N$ incurs minimal error $\epsilon(h)=\mathbb{E}_{(\mathbf{x},\mathbf{y})}[\mathcal{L}(h(x)),y)]$. Among these, $\mathcal{M}$ in Equation \eqref{eq1} is the most crucial factor affecting. However, $L$ and $n(t)$ also influence the performance of the model. The increased $L$ presents a twofold outcome: the provision of augmented information and the challenge of model's adaptability to memory and deal longer length. 

\section{The Proposed MAMCA Method}\label{sec3}

\subsection{Selective State Space Model}\label{subsec3_1}

The State Space Model (SSM) utilizes first-order differential equations to represent the evolution of latent states within a continuous system and employs another set of equation to describe the relationship between its latent states and the output sequence. Given the system's known conditions, the input sequence $x(t)\in\mathbb{R}^D$ can be mapped to the output sequence $y(t)\in\mathbb{R}^N$ through the latent state $h(t)\in\mathbb{R}^N$, and the next state of the system can be predicted as well: 
\begin{equation}
  \begin{aligned}h^{\prime}(t)&=\boldsymbol{A}h(t)+\boldsymbol{B}x(t)\\y(t)&=\boldsymbol{C}h(t)\end{aligned}
\end{equation}where $\boldsymbol{A}\in\mathbb{R}^{N \times N}$ and $\boldsymbol{B}, \boldsymbol{C}\in\mathbb{R}^{N \times D}$ are the system matrices to be learned. To model discrete sequences, the above functions can be discretized using a time step $\Delta$:
\begin{equation}
  \begin{aligned}&h(n)=\overline{\boldsymbol{A}}h(n-1)+\overline{\boldsymbol{B}}x(n),\\&y(n)=\boldsymbol{C}h(n),\end{aligned}
\end{equation}where $\overline{\boldsymbol{A}}=\exp(\Delta \boldsymbol{A})$ and $\overline{\boldsymbol{B}}=(\Delta \boldsymbol{A})^{-1}(\exp(\Delta \boldsymbol{A})-I)\cdot\Delta\boldsymbol{B}$. Upon transforming the continuous form $(\Delta, \boldsymbol{A}, \boldsymbol{B}, \boldsymbol{C})$ into the discrete form $(\overline{\boldsymbol{A}}, \overline{\boldsymbol{B}}, {\boldsymbol{C}})$, the model can be computed in a linear recurrent manner, thereby enhancing computational efficiency \cite{albert_advances_2021}.

To further enhance the capability of long-range dependency modeling through $h$'s memory of $x$, the HiPPO matrix is proposed to augment $\overline{\boldsymbol{A}}$ by the Structured SSM (S4) \cite{albert_Efficiently_2022}. Simultaneously, employing approximate diagonalization, the matrix $\overline{\boldsymbol{A}}$ can be reduced to a Normal Plus Low-Rank (NPLR) representation with only a couple of normal and low-rank components, significantly mitigating the complexity of the recurrent model from $O(L^2)$ to $O(L)$ in Equation \eqref{eq2}: 
\begin{equation}
  \boldsymbol A_{nk}=-\begin{cases}(2n+1)^{1/2}(2k+1)^{1/2}&\text{if }n>k\\n+1&\text{if }n=k\\0&\text{if }n<k\end{cases}
\end{equation}\begin{equation}
  \begin{aligned}
    {\boldsymbol A} &=\boldsymbol{V}\boldsymbol{\Lambda}\boldsymbol{V}^*-\boldsymbol{P}\boldsymbol{Q}^\top \\ &=\boldsymbol{V}\left(\boldsymbol{\Lambda}-(\boldsymbol{V}^*\boldsymbol{P})(\boldsymbol{V}^*\boldsymbol{Q})^*\right)\boldsymbol{V}^*
  \end{aligned}
\end{equation}where $\boldsymbol{V}\in\mathbb{R}^{N \times N}$ is unitary, $\boldsymbol{\Lambda}$ is diagonal, and $\boldsymbol{P}\in\mathbb{R}^{N \times 1}$ and $\boldsymbol{Q}\in\mathbb{R}^{N \times 1}$ are low-rank matrices.

\begin{algorithm}[t]
  \caption{Selective SSM}\label{algo1}
  \begin{algorithmic}[1]
      \Require{$x: (B$(\textbf{batch})$, L$(\textbf{length})$, $$D$(\textbf{dimension})$)$}
      \Ensure{$y: (B, L, D)$}
      \State{$\boldsymbol{A}: (D, $$N$(\textbf{hidden state size})$)$$ \leftarrow \mathrm{Parameter}$ \{structured state matrix, approximately diagonalized from $D \times N \times N$ to $D \times N$\}}
      \State{$\boldsymbol{B}, \boldsymbol{C}: (B, L, N)$ $\leftarrow$ $\mathrm{Linear}_{N}(x)$, $\mathrm{Linear}_{N}(x)$}
      \State{${\Delta}: (B, L, D)\leftarrow \mathrm{Softplus}(\mathrm{Parameter} + \mathrm{Broadcast}_{D}(\mathrm{Linear}_{1}(x)))$}
      \State{$\overline{\boldsymbol{A}}, \overline{\boldsymbol{B}}: (B, L, D, N)$ $\leftarrow$ $\mathrm{Parameter}$ discretize$({\Delta}, \boldsymbol{A}, \boldsymbol{B})$ \{Input-dependent parameters and discretization\}}
      \State{$y: (B, L, D)$ $\leftarrow$ $\mathrm{SelectiveSSM}(\overline{\boldsymbol{A}}, \overline{\boldsymbol{B}}, \boldsymbol{C})$$(x)$}
\end{algorithmic}
\end{algorithm}

Due to the time-invariance of $(\boldsymbol{A}, \boldsymbol{B}, \boldsymbol{C})$ within the conventional SSM, their performance on processing long-sequence data is suboptimal. A novel form of Selective SSM (Mamba) \cite{albert_mamba_2023} introduces selectivity by dynamically adjusting its states based on the current input, selectively propagating or ignoring information. As illustrated in Algorithm \ref{algo1}, Mamba block incorporates the length of the input sequence and batch size, rendering $\boldsymbol{B}$, $\boldsymbol{C}$ and $\Delta$, input-dependent. Following discretization, the $\overline{\boldsymbol{A}}=\exp(\Delta \boldsymbol{A})$ allowing A to correlate with the inputs. For instance, a larger $\Delta$ biases the model towards prioritizing current input over previous information; in contrast, a smaller $\Delta$ retains more historical data.

Given that these parameters vary according to the input, the system is no longer linear and time-invariant. Mamba introduces a hardware-aware algorithm: it reads $(\Delta, \boldsymbol{A}, \boldsymbol{B}, \boldsymbol{C})$ directly into Static Random Access Memory (SRAM) from the High Bandwidth Memory (HBM) before computing $\overline{\boldsymbol{A}}$ and $\overline{\boldsymbol{B}}$. Subsequent discrete processing and parallel scanning within SRAM, followed by summation with ${\boldsymbol{C}}$, yields the resulting output that is then returned to HBM. Compared to the hardware data flow of other models, Mamba reduces HBM-SRAM interactions by one during once propagation, with a total read/write volume of $O(BLN)$, decreasing by $O(N)$.

\vspace{-.3cm}
\subsection{Soft-thresholding Denoising}\label{subsec3_2}

\begin{figure*}[htbp]
  \centering
  {\includegraphics[width=2\columnwidth]{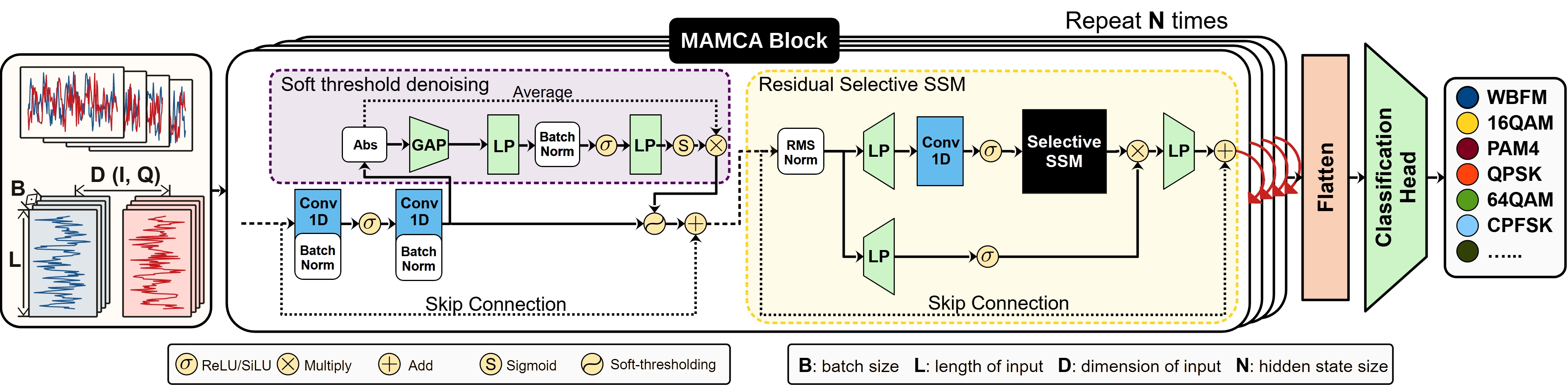}}
  \caption{The structure of proposed MAMCA framework. The MAMCA block consist of a denoising unit and a Selective SSM unit.}\label{framework}
  \vspace{-.4cm}
\end{figure*}

Facing low SNR signals, the unique characteristics of modulation schemes are difficult to extract due to noise interference, resulting in a decline in the learning capability of AMC models. Soft-thresholding is a key method for signal denoising, which attenuates noise features by eliminating the influence of specific range characteristics \cite{zhao_deep_2020}. The soft-thresholding function is as follows:
\begin{equation}
  \begin{aligned}
    y=\begin{cases}x+\tau&\text{if }x<-\tau\\0&\text{if }-\tau\leq x\leq\tau\\x-\tau&\text{if }x>\tau\end{cases}
  \end{aligned}
\end{equation}where $x$ represents the input feature, $y$ represents the output feature, and $\tau$ is the threshold parameter. Soft-thresholding converts features close to zero into zero, while preserving useful negative features. Its derivative is either 0 or 1, which can effectively prevent the problems of gradient vanishing and gradient explosion, as shown in the following equation:
\begin{equation}
  \begin{aligned}
    \frac{\partial y}{\partial x}=\begin{cases}1&\text{if }x<-\tau \text{ or } x>\tau\\
      0&\text{if }-\tau\leq x\leq\tau \end{cases}
  \end{aligned}
\end{equation}

\vspace{-.3cm}
\subsection{MAMCA framework}\label{subsec3_3}

The proposed MAMCA framework is presented in Fig. \ref{framework}. The input form consists of I/Q sampled signals. MAMCA is primarily composed of two core modules: MAMCA blocks, and a classification head. 

Among these, the MAMCA block perform feature extraction on the original input signal. As previously mentioned, we aspire to address the issue of noise interference by soft-thresholding. However, manually setting a suitable fixed threshold is challenging. To resolve this issue, we refer to the Deep Residual Shrinkage Network \cite{zhao_deep_2020}, which employs attention mechanism to automatically learn a reasonable threshold. We introduce a special attention module to estimate the threshold, in which the feature map undergoes absolute value processing and global average pooling, followed by a linear projection to obtain a scaling parameter. The sigmoid function then compresses these scaling parameters within the $[0, 1]$ range. Finally, the scaling parameter multiplies the average result of the absolute values of the feature map to derive the soft threshold. By setting up such a structure, we guarantee the soft threshold positive, while also restricting its size to remain smaller than the maximum value of the feature map, thus serving the purpose of filtering and retaining. Subjecting the original input to soft-thresholding preserves the main attributes, while those deemed as noise are nullified. This enhances the prominence of high-level features. The Selective SSM module is the main component, meeting the needs for long-term dependency, efficient sequential memory occupancy and high execution speed. MAMCA ultimately completes the AMC task through a classification head.

\section{Experiment Results and Analysis}\label{sec4}

\subsection{Implementation Details}\label{subsec5_1}
\vspace{-.4cm}
\begin{table}[htbp]
  \centering
  \caption{Detail Parameters of Datasets}
    \begin{tabular}{cccc}
      \toprule
       \textbf{Parameter} & \textbf{RML2016.10a} & \textbf{RML2016.10b} & \textbf{TorchSig-QAM} \\
      \midrule
    \makecell{Modulation\\number} & 11 & 10 & 6 \\
    \makecell{Samples per\\modulation} & 20000 & 20000 & 3072 \\
    Signal length & 128   & 128   & \makecell{128, 256, 512, \\1024, 2048, 4096} \\
    SNR range & -20:2:20dB & -20:2:20dB & -15:5:20dB \\
      \bottomrule
    \end{tabular}%
  \label{dataset}%
\end{table}%

The RML2016 \cite{Timothy_radio_2016} dataset is widely utilized for AMC. The RML2016.10b subset encompasses WBFM, QPSK, 64QAM, 16QAM, PAM4, GFSK, CPFSK, BPSK, AM-DSB, and 8PSK, while RML2016.10a additionally includes AM-SSB. Furthermore, we have also selected the ideal QAM family dataset from the open-source TorchSig \cite{luke_large_2022} dataset, which comprises 64QAM, 256QAM, 1024QAM, 32QAM-Cross, 128QAM-Cross, and 512QAM-Cross. For this ideal dataset, we used the HackRF One for transmission and the URSP B210 for reception. The training-testing ratio for all datasets is 8:2.

To further appraise the performance of our model, we employed several AMC benchmarks, including PET-CGDNN \cite{zhang_Efficient_2021}, MCLDNN \cite{xu_Spatiotemporal_2020}, ICAMC \cite{Hermawan_CNN_2020}, and DAE \cite{ke_real_2022}. In addition, comparisons were made with 1D-Resnet, Vallina-Transformer, and Complex-Valued CNN (CVCNN). All models are built on Python 3.8 in PyTorch 2.0.1 and trained on an NVIDIA GeForce GTX 3080 Ti GPU with Adam optimizer.

\vspace{-.3cm}
\subsection{Experiment Results}\label{subsec5_2}

\subsubsection{ Accuracy Comparison}\label{subsec5_2_1}

\begin{figure*}[t]
  \centering
  \subfloat[RML2016.10a]{\includegraphics[width=.66\columnwidth]{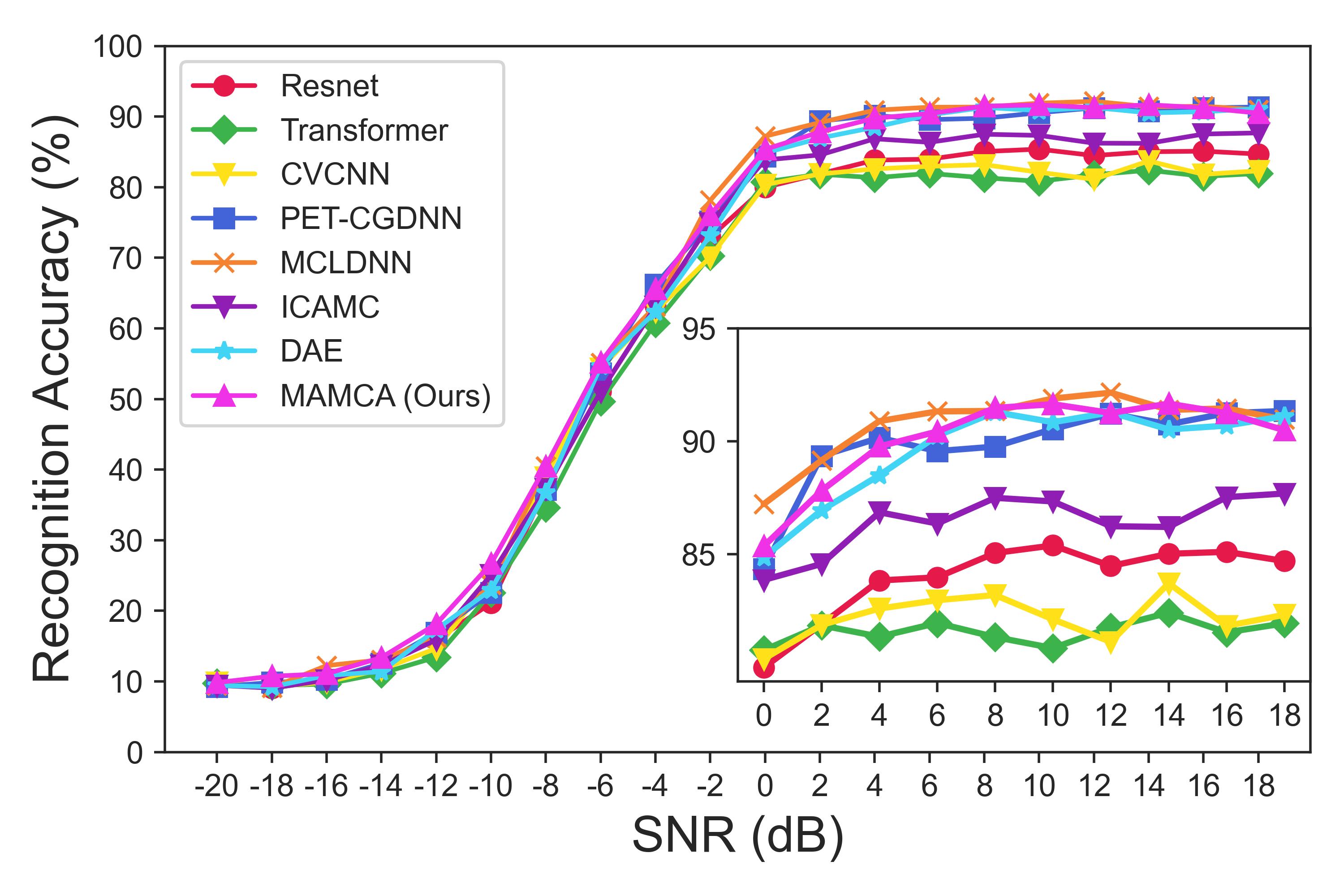}\label{201610a}}
  \subfloat[RML2016.10b]{\includegraphics[width=.66\columnwidth]{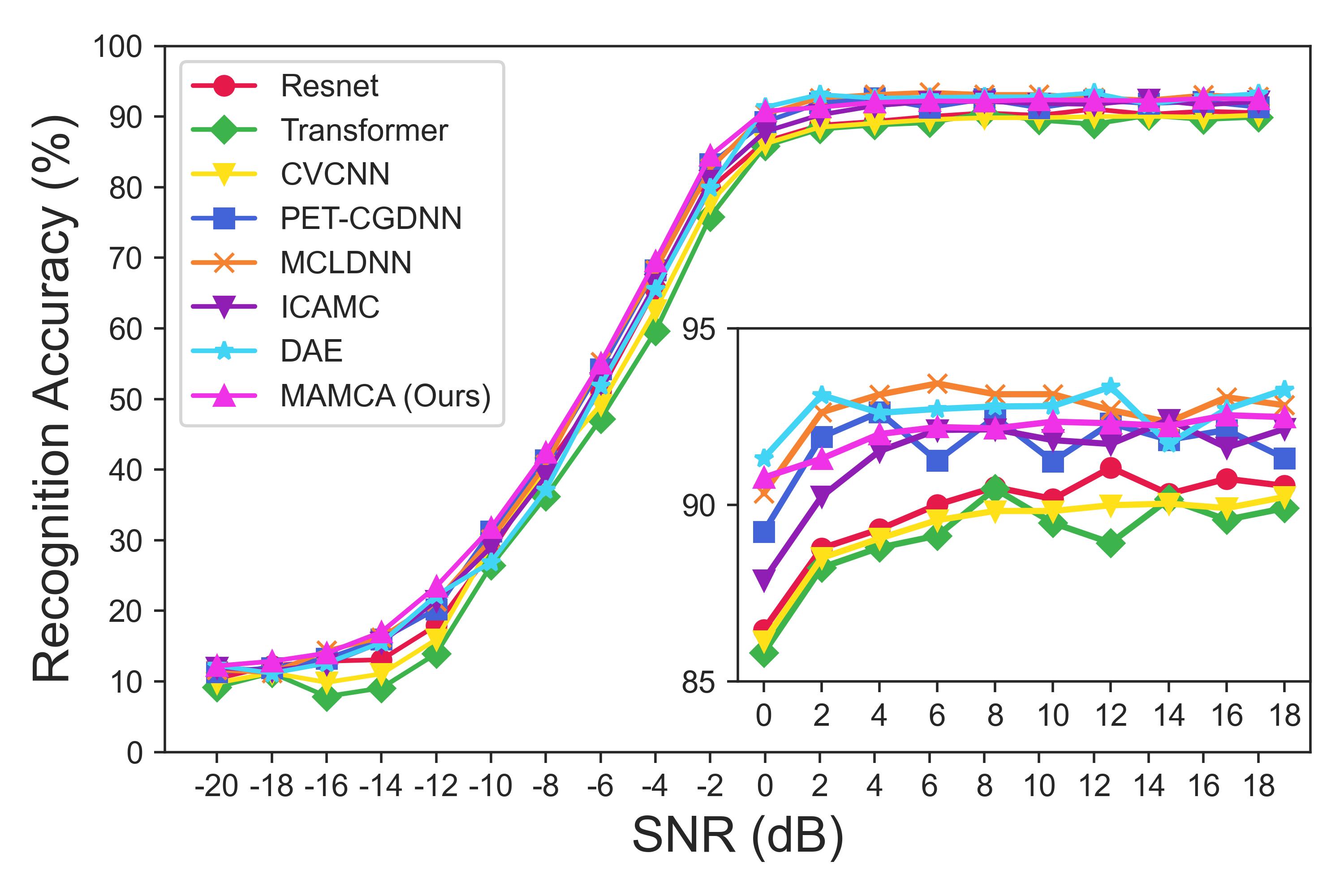}\label{201610b}}
  \subfloat[TorchSig-QAM]{\includegraphics[width=.66\columnwidth]{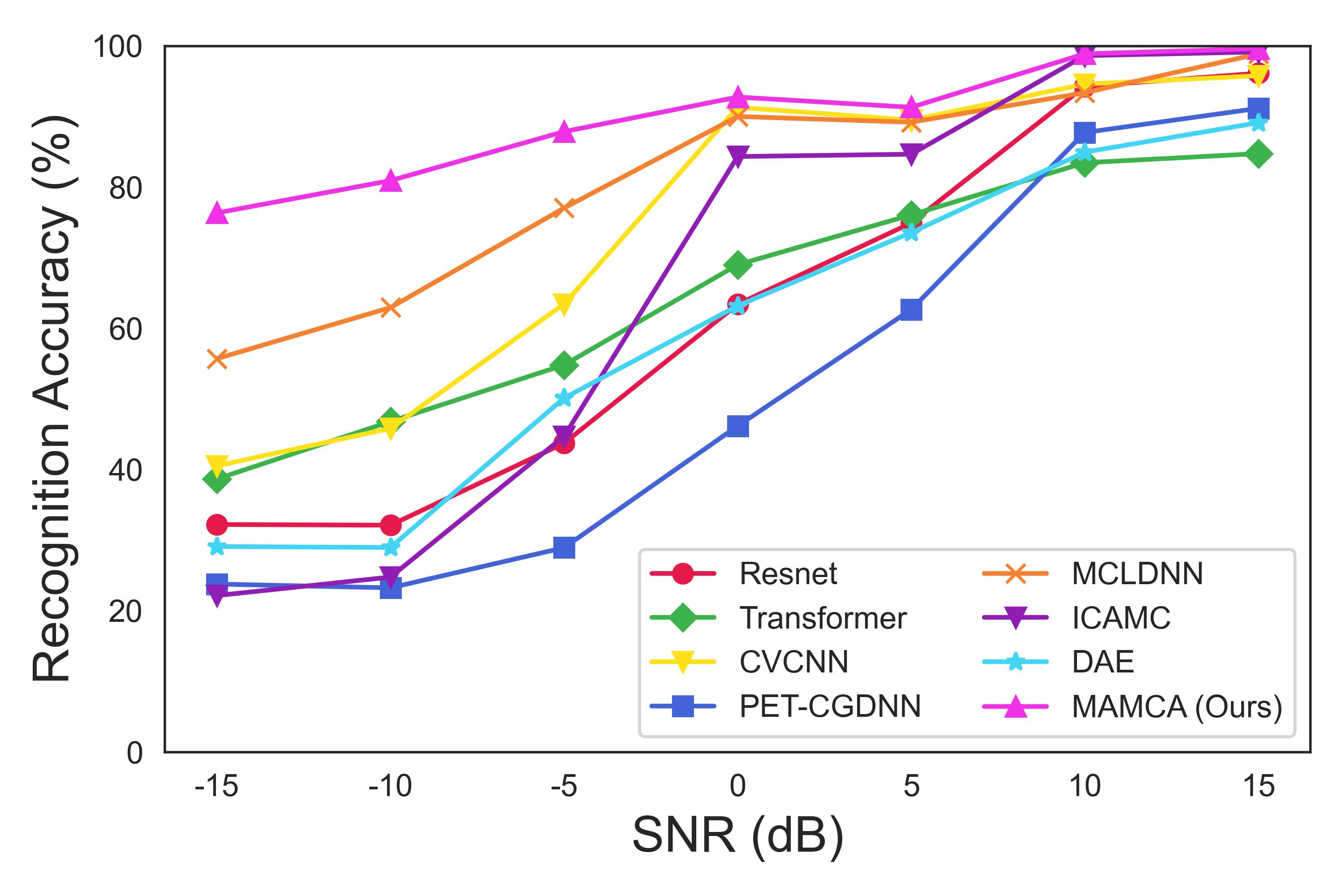}}
  \caption{Recognition accuracy of MAMCA and other AMC models on the RML2016.10a, RML2016.10b and TorchSig-QAM dataset.}\label{RML}
  \vspace{-.4cm}
\end{figure*}

\begin{figure*}[t]
  \centering
  \subfloat[Effect of length on recognition accuracy]{\includegraphics[width=.66\columnwidth]{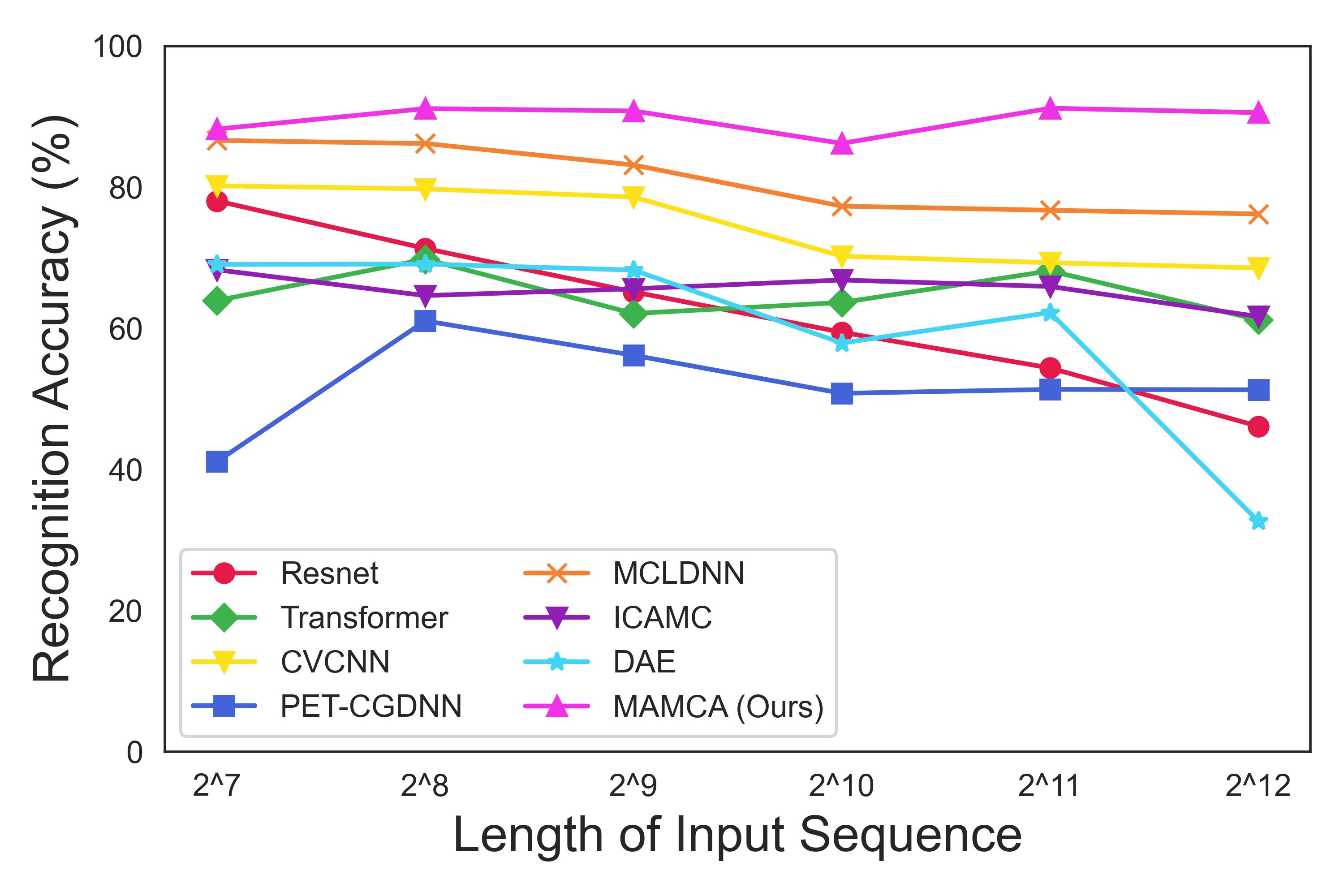}\label{acc_length}}
  \subfloat[Effect of length on training time]{\includegraphics[width=.66\columnwidth]{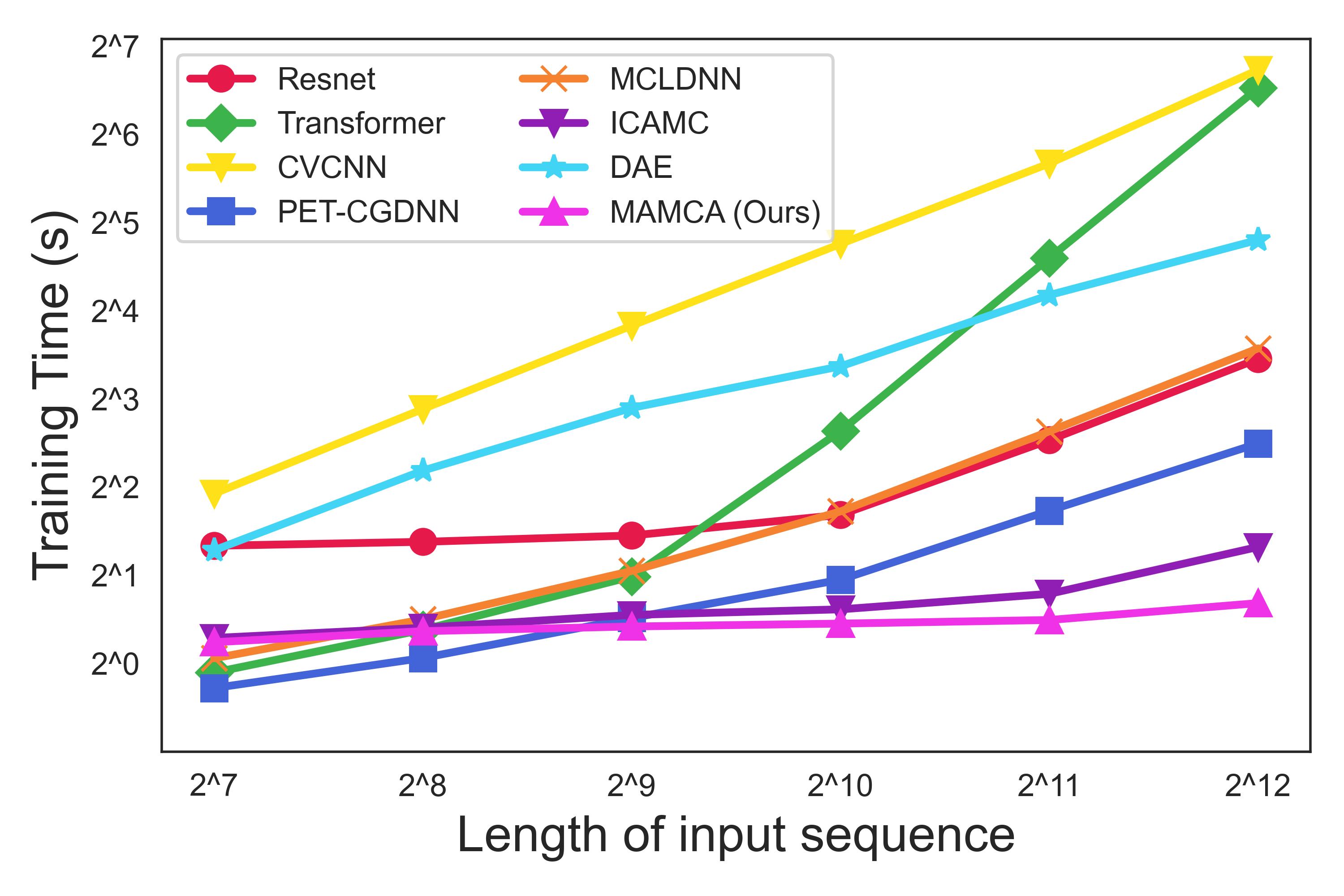}\label{train_time_length}}
  \subfloat[Effect of length on inference time]{\includegraphics[width=.66\columnwidth]{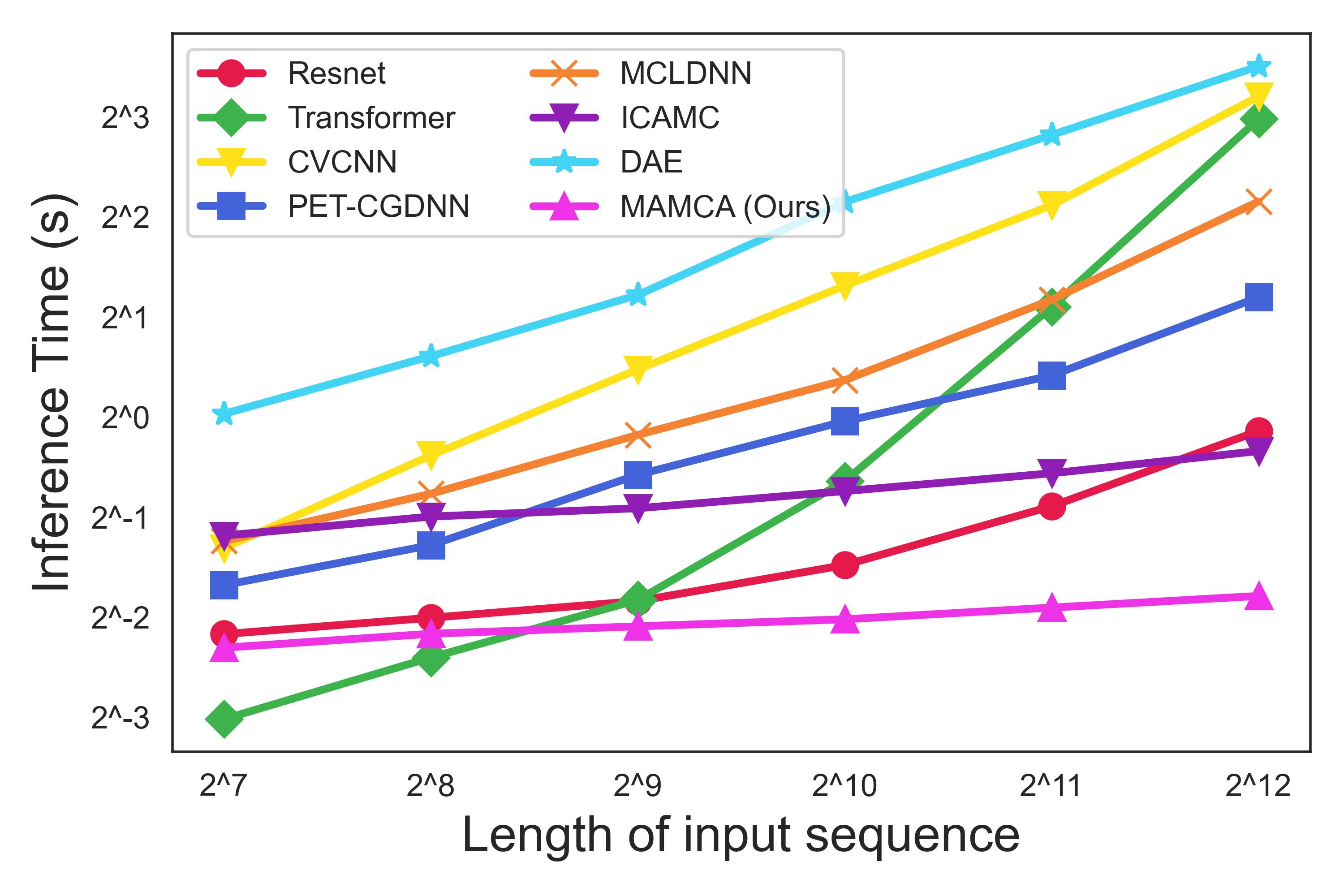}\label{inference_time_length}}
  \caption{Recognition accuracy, training time and inference time of MAMCA and other AMC models influenced by extending sample lengths.}\label{length_effect}
  \vspace{-.4cm}
\end{figure*}


As shown is Fig. \ref{RML}, we conducted a comparative analysis of MAMCA alongside other AMC models across three distinct datasets. Within the confines of the RML2016 dataset, characterized by a concise sample length of 128, MAMCA manifests strong competitiveness. Moreover, when tested on the TorchSig-QAM dataset encompassing a length spectrum from 128 to 4096, MAMCA's performance is notably superior. In environments with various SNR, MAMCA achieves a recognition rate that transcends that of competing models. The incorporation of the denoising units renders MAMCA relatively impervious to the adverse effects present in low SNR recognition tasks, securing an accuracy above 75\% even at -15dB, thereby outperforming the second-best by approximately 20\%. Details in Table \ref{acc} shows the average accuracy integrated across all SNR scenarios. MAMCA secures a marked advantage, which is particularly evident when contending with the TorchSig-QAM dataset, where increased noise levels and subtle distinctions between modulation schemes are prevalent. MAMCA's predominance ranges from 8.65\% to a substantial 37.71\% over other models.

\begin{table}[t]
  \centering
  \caption{Comparison of Accuracy of Different Models}
    \begin{tabular}{lccc}
      \toprule
    \multicolumn{1}{c}{\textbf{Model}} & \textbf{RML2016.10a} & \textbf{RML2016.10b} & \textbf{TorchSig-QAM} \\
      \midrule
    Resnet & 57.10  & 61.48  &  62.41 \\
    Transformer & 55.34  & 59.60  & 74.43  \\
    CVCNN & 56.37  & 60.51  & 64.80  \\
    PET-CGDNN\cite{zhang_Efficient_2021} & 60.58  & 63.37  & 51.96  \\
    MCLDNN\cite{xu_Spatiotemporal_2020} & \textbf{61.42}  & \underline{63.92}  & \underline{81.02}  \\
    ICAMC\cite{Hermawan_CNN_2020} & 58.62  & 62.70  & 65.49  \\
    DAE\cite{ke_real_2022}   & 60.22  & 63.17  & 59.64  \\
    MAMCA (Ours) & \underline{60.79}  & \textbf{64.05}  & \textbf{89.67}  \\
      \bottomrule
    \end{tabular}%
  \label{acc}%
  \vspace{-.4cm}
\end{table}%

The MAMCA also evidences its superiority in memory occupancy for signals of extended length. As depicted in Fig. \ref{acc_length}, when confronted with signal length of 128 consistent with the RML2016 dataset parameters, the accuracy of competing models hovers above 60\%. However, this trend undergoes a discernible shift as the sample data length swells to 4096, as featured in the TorchSig-QAM dataset. Under this condition, other models demonstrate a spectrum of performance drop-offs. In stark contrast, MAMCA sustains a steady accuracy, regardless of signal length variations. Notably, MAMCA consistently outperforms all other models across the full of data lengths, particularly at the upper end with signal length at 4096, where it boasts an accuracy margin approximately 15\% greater than the next best.

\subsubsection{ Efficiency Comparison}\label{subsec5_2_2}

In addition to its outstanding accuracy, MAMCA also boasts an impressively light scale and quick processing speed, as depicted in Fig. \ref{train_time_length} and \ref{inference_time_length}. Within the range from a length of 128 to 4096, both the training and inference time for MAMCA are exceedingly low. Moreover, as the points of sample increase, the time consumption of MAMCA remains almost constant, with only a slight increase, while other models exhibit a high linear or even quadratic growth in duration. Coupled with MAMCA's stability in accuracy on long samples as discussed earlier, this suggests that MAMCA is capable of adapting to enormous training and predicting throughput.

\begin{figure}[t]
  \centering
  {\includegraphics[width=1\columnwidth]{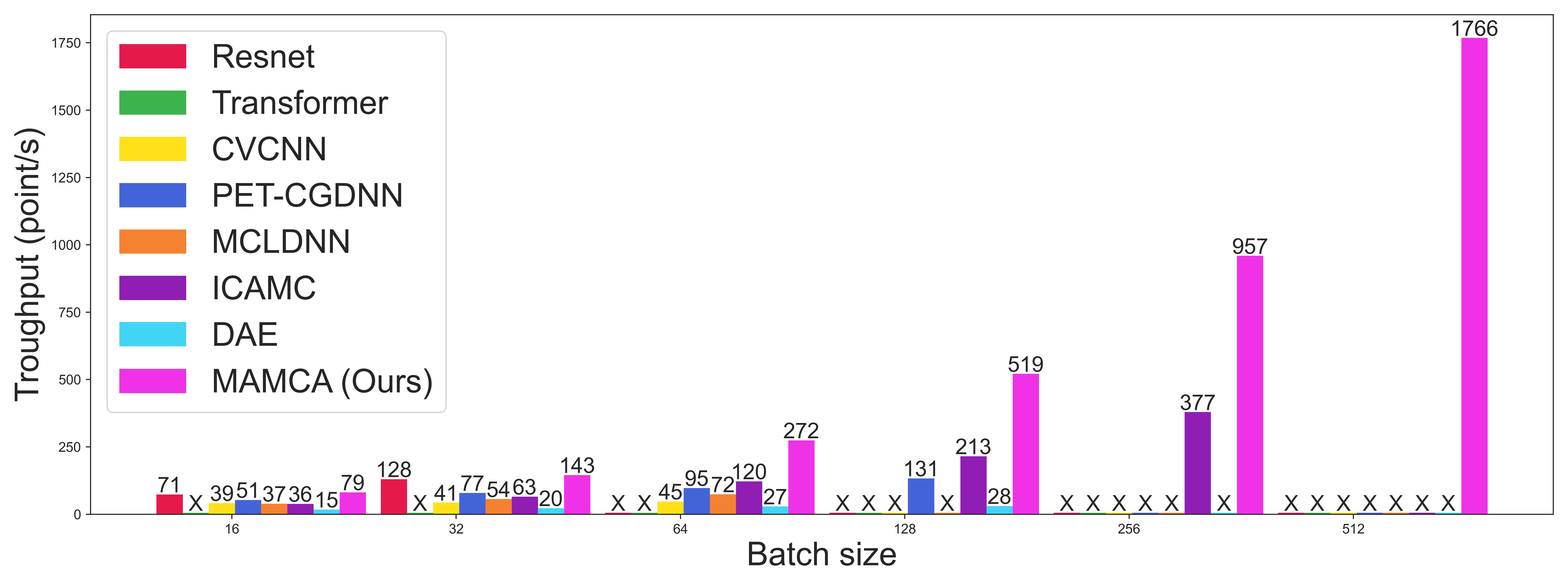}}
  \vspace{-.6cm}
  \caption{Throughput comparison of length 4096 when batch size grows. `X' represent that the GPU (10240Mb) is out of memory.}\label{throughput}
  \vspace{-.3cm}
\end{figure}

\begin{table}[t]
  \centering
  \caption{Comparison of Complexity of Different Models}
    \begin{tabular}{lllll}
      \toprule
      \makecell[c]{\textbf{Model}} & \multicolumn{1}{c}{\textbf{FLOPS/M}} & \multicolumn{1}{c}{\textbf{Params/M}} & \multicolumn{1}{c}{\textbf{Train}} & \multicolumn{1}{c}{\textbf{Inference}} \\
    \midrule
    Resnet & 1.12E+04 & 5.38E+01 & 1.10E+01 & 9.09E-01 \\
    Transformer & 4.08E+02 & 3.84E-02 & 9.21E+01 & 7.89E+00 \\
    CVCNN & 3.10E+03 & 6.81E+01 & 1.06E+02 & 9.22E+00 \\
    PET-CGDNN & 8.69E+01 & 7.92E+01 & 5.64E+00 & 2.29E+00 \\
    MCLDNN & 1.17E+03 & 4.06E+02 & 1.19E+01 & 4.45E+00 \\
    ICAMC & 9.50E+02 & 3.38E+01 & \underline{2.50E+00} & \underline{7.91E-01} \\
    DAE   & \textbf{5.36E-01} & \textbf{1.49E-01} & 2.79E+01 & 1.14E+01 \\
    MAMCA (Ours) & \underline{7.39E+01} & \underline{1.68E+01} & \textbf{1.61E+00} & \textbf{2.46E-01} \\
    \bottomrule
    \end{tabular}\label{efficiency}%
    \vspace{-.4cm}
\end{table}

Table \ref{efficiency} compares different models on the number of flops, storage overhead, training time per epoch, and inference time per test set for samples with a length of 4096. MAMCA exhibits exceptional efficiency, with all metrics at the lowest magnitude. The training duration of MAMCA is 1.52\% to 64.52\% times that of other models, while the inference duration is 0.27\% to 31.06\% times that of other models. This implies that MAMCA has more extensive deployment conditions.

In addition, the GPU occupancy of MAMCA is incredibly small, which when combined with its rapid inference capability, enables it to achieve an astonishing throughput on longer length data. As shown in Fig \ref{throughput}, as the batch size increases, the memory usage of MAMCA rises gradually, thereby allowing it to process a larger number of long sequences concurrently.

\subsubsection{ Ablation Study}\label{subsec5_2_3}

To further verify the effectiveness of each module in MAMCA, we conducted an ablation study. As shown in Table \ref{ablation}, the average accuracy of the model on various datasets decreases when only retaining either the denoising unit or Selective SSM. This indicates that these modules both play a positive role when combined.

\begin{table}[h]
  \centering
  \caption{Ablation Study on RSBU and Selective SSM Module}
    \begin{tabular}{llll}
      \toprule
      \multicolumn{1}{c}{\textbf{Model}} & \textbf{RML2016.10a} & \textbf{RML2016.10b} & \textbf{TorchSig-QAM} \\
      \midrule
    \makecell{Delete\\Selective SSM} & 47.61 ($\downarrow 13.18$) & 52.92 ($\downarrow 11.13$) & 75.81 ($\downarrow 13.86$) \\
    \makecell{Delete\\denoising unit} & 43.35 ($\downarrow 17.44$) & 50.23 ($\downarrow 13.82$) & 66.24 ($\downarrow 23.43$) \\
      \bottomrule
    \end{tabular}%
  \label{ablation}%
\end{table}%

\section{Conclusion}\label{sec5}

In this letter, we propose a novel MAMCA model for AMC that enhances the accuracy and efficiency with extended signal length by incorporating Selective SSM and denoising units. Experimental results on multiple datasets demonstrate that MAMCA, in comparison to existing models, not only possesses a highly competitive accuracy rate but also benefits from significantly lower temporal and spatial overheads, thus obtaining large throughput. Moreover, MAMCA exhibits improved robustness over longer sequences. These indicate that MAMCA has a higher application value across a broader range of IoT scenarios. Future work includes better utilization of the SSM within AMC to further enhance accuracy, timeliness, and robustness over extended signal durations.
\vspace{-.3cm}

\bibliographystyle{IEEEtran}
\bibliography{ref}

\end{document}